\documentclass[twocolumn,english,prl,floatfix,showpacs]{revtex4-1}
\usepackage[T1]{fontenc}
\usepackage[latin9]{inputenc}
\usepackage{amsmath}
\usepackage{amssymb}
\usepackage{graphicx}
\usepackage{wasysym}
\usepackage{caption}
\usepackage{subcaption}

\makeatletter

\graphicspath{ {./figures/} }

\usepackage{babel}

\@ifundefined{showcaptionsetup}{}{%
 \PassOptionsToPackage{caption=false}{subfig}}
\usepackage{subfig}
\makeatother

\usepackage{babel}
\begin{document}

\title{Continuous Time Quantum Walks in finite Dimensions }

\author{Shanshan Li and Stefan Boettcher}

\affiliation{Department of Physics, Emory University, Atlanta, GA 30322; USA }
\begin{abstract}
We consider the quantum search problem with a continuous time quantum
walk for networks of finite spectral dimension $d_{s}$ of the network
Laplacian. For general networks of fractal (integer or non-integer)
dimension $d_{f}$, for which in general $d_{f}\not=d_{s}$, it suggests
that $d_{s}$ is the scaling exponent that determines the computational
complexity of the search. Our results are consistent with those of
Childs and Goldstone {[}Phys. Rev. A 70 (2004), 022314{]} for lattices
of integer dimension, where $d=d_{f}=d_{s}$. For general fractals,
we find that the Grover limit of quantum search can be obtained whenever
$d_{s}>4$. This complements the recent discussion of mean-field (i.e.,
$d_{s}\to\infty$) networks by Chakraborty et al. {[}Phys. Rev. Lett.
116 (2016), 100501{]} showing that for all those networks spatial
search by quantum walk is optimal. 
\end{abstract}


\maketitle
Quantum walks present one of the frameworks for which quantum computing
can satisfy its promise to provide a significant speed-up over classical
computation. Grover \cite{Gro97a} has shown that a quantum walk can
locate an entry in an unordered list of $N$ sites in a time that
scales as $\sim\sqrt{N}$, a quadratic speed-up over classical search
algorithms. However, that finding was based on a list in which all
sites are interconnected with each other, thus, raising the question
regarding the impact of geometry on this result. Note, for instance,
that if the walk had to pass the list over a linear, \emph{1d}-line
of sites, no quantum effect would provide an advantage over simply
passing every site until the desired entry is located. And for obvious
technical reasons, the design of a quantum algorithm that could satisfy
the Grover limit especially for lists embedded in \emph{2d}-space
is particularly desirable. 

A continuous time quantum walk (CTQW) in any geometry can be defined
by the Hamiltonian
\begin{eqnarray}
\mathcal{H} & = & \gamma\thinspace\mathcal{L}-\left|w\right\rangle \left\langle w\right|,\label{eq:Hamiltonian}
\end{eqnarray}
where $\mathcal{L}$ is the Laplacian matrix, and $\left|w\right\rangle \left\langle w\right|$
is the projection operator (the oracle) for some target site $w$.
With that, a quadratic speedup for quantum search has been shown for
high dimensional graphs such as the complete graph\ \cite{farhi1998QWcompleteG}
and the hypercube\ \cite{childs2002hypercube,farhi2000hypercube}.
Recently, it has been proven also to be optimal on Erdös-Renyi graphs
with $N$ sites as long as the existence probability for an edge between
any two sites is $p\geq\log^{3/2}N/N$ or the graphs are regular,
i.e., they have the same degree for every site\ \cite{chakraborty2015randomG}.
However, Childs and Goldstone \cite{Childs04} have shown that such
a quantum search can reach the Grover limit on lattices in dimensions
$d>4$ only, while in $d=4$, the running time to achieve a success
probability of order $1$ is $O\left(\sqrt{N}\log^{3/2}N\right)$,
with increasing deviations from $\sqrt{N}$-scaling for $d=3$ and
2. In contrast, a discrete-time, coined version of a quantum walk
has been proposed by which quantum search falls short of the Grover
limit in $d=2$ only by logarithmic factors \cite{AKR05,PortugalBook}.
A better understanding of this discrepancy, its origin and potential
remedies, is of considerable interest. 

Here, we generalize those results to arbitrary real (fractal) dimensions
$d$. Fractals generally possess both a fractal dimension $d_{f}$
and a spectral dimension $d_{s}$ that can vary independently to characterize
their geometry \cite{Alexander82}, while for regular lattices $d_{f}=d_{s}=d$.
Scaling as well as exact renormalization group (RG) arguments show
that the spectral dimension $d_{s}$ of the lattice Laplacian controls
the ability of CTQW to saturate the Grover limit generally, even for
cases where $d_{f}\not=d_{s}$. 

Certain graphs with fractal dimensions have been considered previously
for search with CTQW \cite{Agliari2011}. The fractals chosen there
include dual Sierpinski gaskets, T-fractals, Cayley trees, and Cartesian
products between Euclidean lattices and dual Sierpinski gaskets, with
a variety of fractal and spectral dimensions. Based on numerical simulations,
it was suggested that whether CTQW provides quadratic speedup is determined
together by a spectral dimension larger than 4 and by the overlaps
of the initial state with the ground and first excited state of the
Hamiltonian. These overlaps undergo a critical transition near the
closest ``gap'' between both levels, controlled by the choice of
$\gamma$ in Eq. (\ref{eq:Hamiltonian}). In this paper, we explicitly
relate the transition in $\gamma$ to (derivatives of) the Laplacian
determinant using a spectral $\zeta$-function. Using exact RG \cite{BoLi15},
we have shown elsewhere that the asymptotic scaling of the Laplacian
determinant is described uniquely in terms of $d_{s}$. 

The continuous time quantum walk on a graph is determined by the Schrödinger
equation evolving in a Hilbert space spanned by the $N$-position
site-basis $\left|x\right\rangle $, 

\begin{eqnarray}
i\thinspace\frac{d\Psi_{x}(t)}{dt} & = & \sum_{y}\mathcal{H}_{xy}\Psi_{x}(t),\label{eq:SchrodingerEq}
\end{eqnarray}
where $\Psi_{x}(t)=\left\langle x|\Psi(t)\right\rangle $ is the complex
amplitude at site $x$, and $\mathcal{H}$ is the Hamiltonian defined
in Eq. (\ref{eq:Hamiltonian}). The search typically evolves from
an initial state that is prepared as the uniform superposition over
all sites \cite{Gro97a}, $\left|s\right\rangle =\frac{1}{N}{\displaystyle \sum_{x}\left|x\right\rangle }$.
The complete graph is a special case of CTQW, where it suffices to
consider the subspace spanned by $\left|s\right\rangle $ and $\left|w\right\rangle $
on which the Hamiltonian acts nontrivially. At $\gamma N=1$, the
ground and first excited state are respectively $\left(\left|w\right\rangle \pm\left|s\right\rangle \right)/\sqrt{2}$
with a energy gap of $2/\sqrt{N}$. The search Hamiltonian achieves
success by driving the system from state $\left|s\right\rangle $
to $\left|w\right\rangle $ with a transition probability $\varPi_{s,w}=\left|\left\langle w\left|e^{-iHt}\right|s\right\rangle \right|^{2}=\sin\left(t/\sqrt{N}\right)$
that reaches unity first at time $t=\frac{\pi}{2}\sqrt{N}$. For a
general geometry, the ground and first excited state are more complicate
than a superposition of $\left|s\right\rangle $ and $\left|w\right\rangle $.
Yet, the objective of CTQW remains two-fold: (1) find a critical value
$\gamma=\gamma_{c}$ such that the overlaps between $\left|s\right\rangle $
as well as $\left|w\right\rangle $ and the ground and first excited
state are substantial, and (2) ascertain that at this critical point
the Hamiltonian drives a transition from $\left|s\right\rangle $
to $\left|w\right\rangle $ in a time $t\sim1/\left(E_{1}-E_{0}\right)\sim\sqrt{N}$.

\begin{figure*}
\centering
\begin{subfigure}{0.27\textwidth}
\centering
\includegraphics[bb=0bp 200bp 558bp 805bp,clip,width=\textwidth]{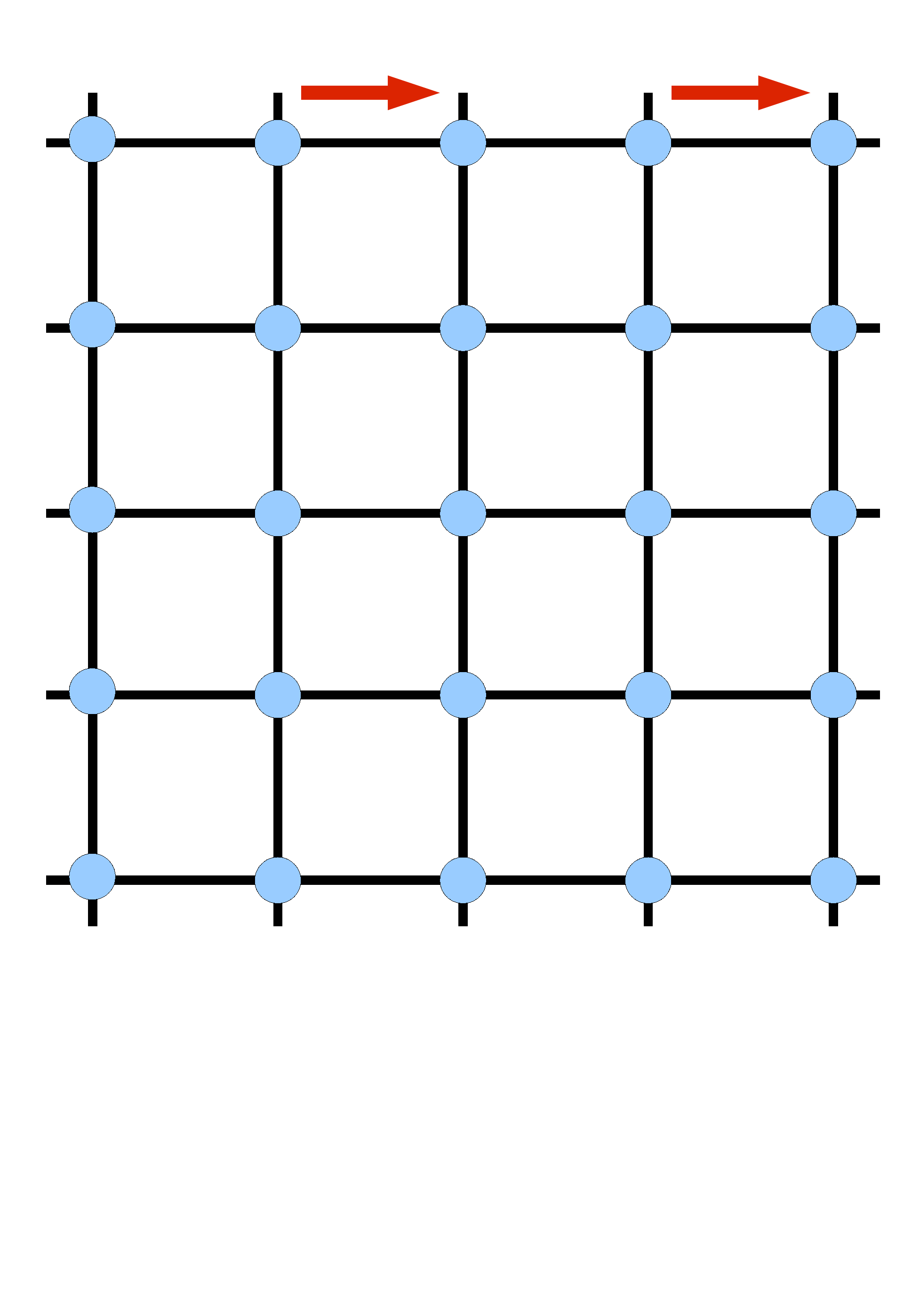}
\caption{}
\end{subfigure}
\begin{subfigure}{0.27\textwidth}
\centering
\includegraphics[bb=0bp 200bp 558bp 805bp,clip,width=\textwidth]{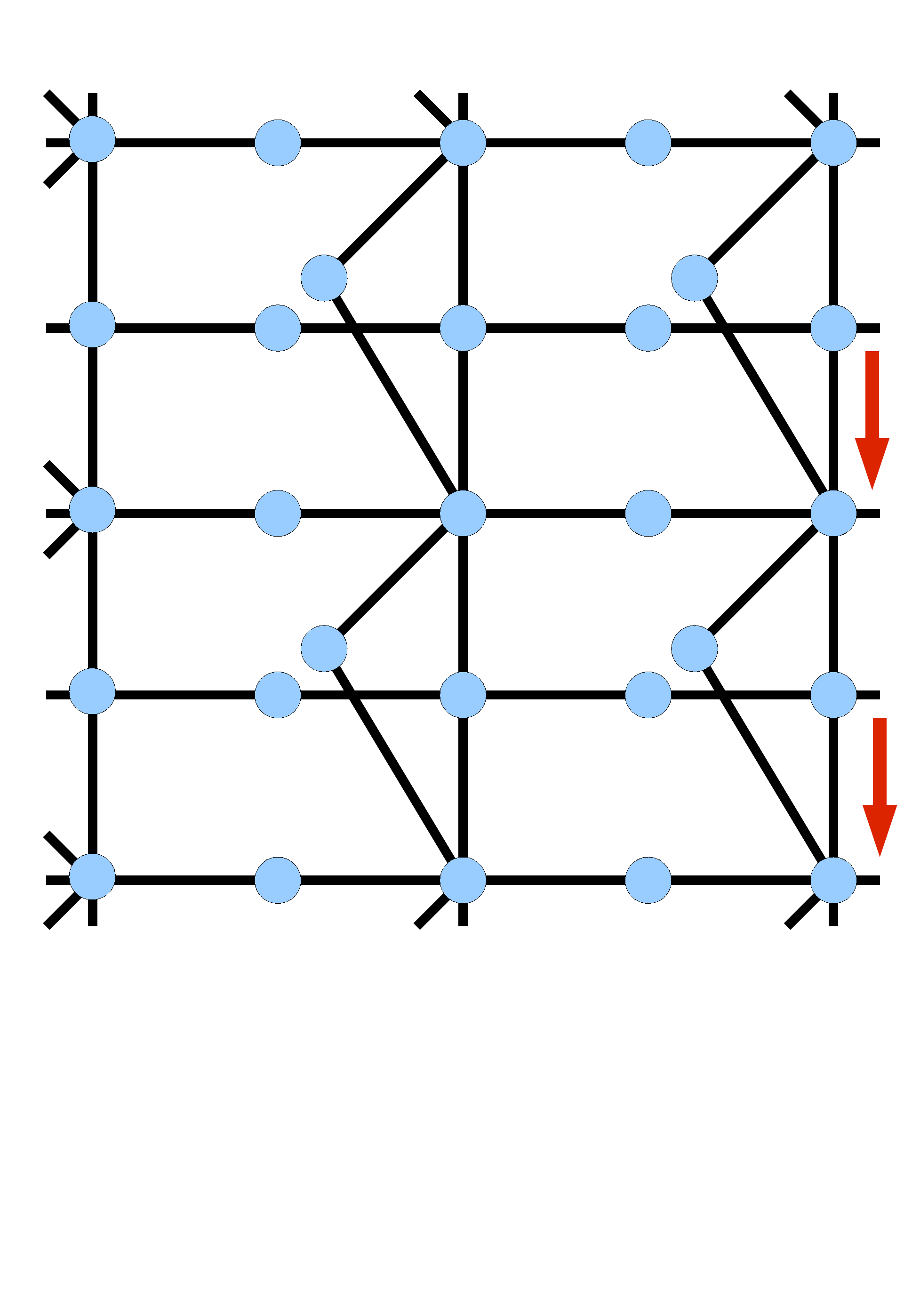}
\caption{}
\end{subfigure}
\begin{subfigure}{0.27\textwidth}
\centering
\includegraphics[bb=0bp 200bp 558bp 805bp,clip,width=\textwidth]{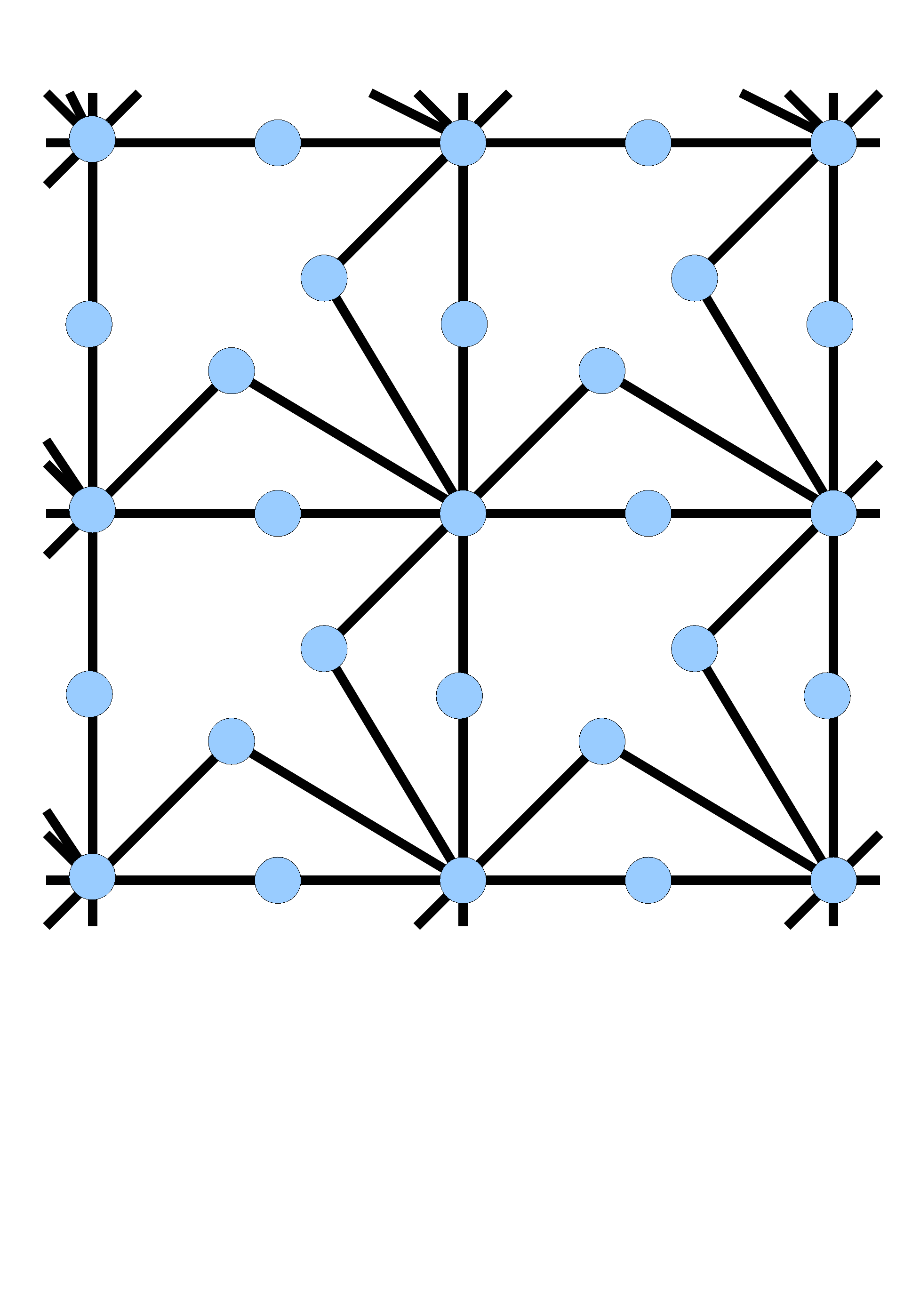}
\caption{}
\end{subfigure}
\begin{subfigure}{0.16\textwidth}
\centering
\includegraphics[bb=70bp 240bp 250bp 580bp,clip,width=\textwidth]{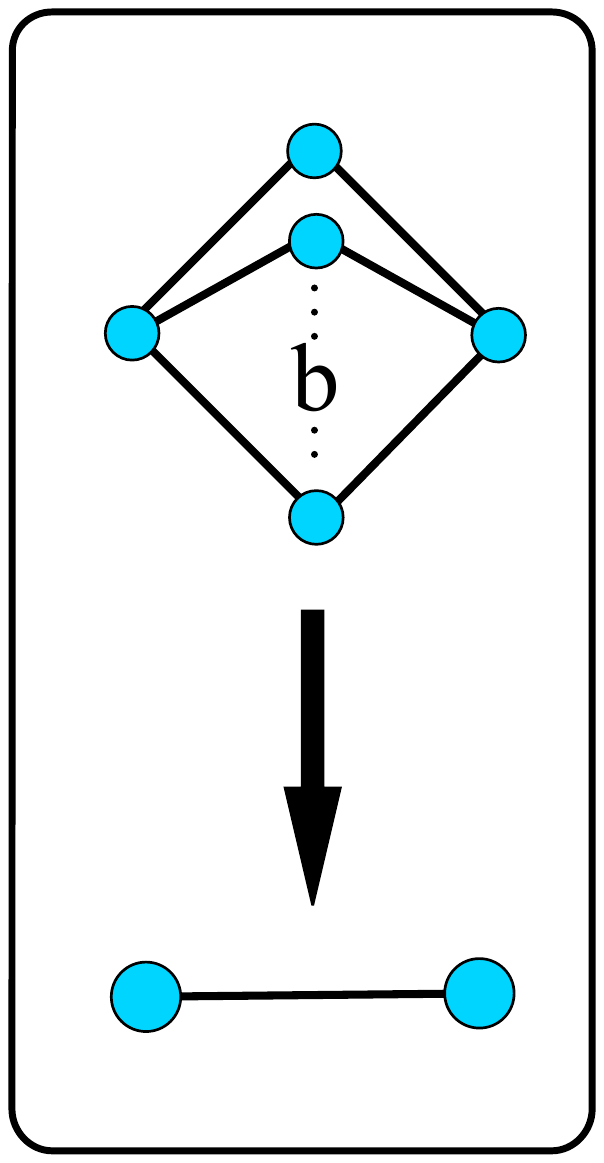}
\caption{}
\end{subfigure}
\captionsetup{justification=raggedright, singlelinecheck=false}
\caption{\label{fig: MKlattice} Bond-moving scheme in the Migdal\textendash Kadanoff
renormalization group (MKRG) \cite{Migdal76,Kadanoff76}, here for
a square lattice ($d=2$) with rescaling length $l=2$ and branching
number $b=2$. Starting from the lattice with unit bonds (a), bonds
in intervening hyper-planes are projected onto every $l^{{\rm th}}$
plane in one direction (b), then subsequent directions (c), to re-obtain
a similar hyper-cubic lattice of bond-length $l$, as in (a). The
renormalized bonds in this case consist of $b=2$ branches, each of
a series of $l=2$ bonds; the general RG-step for $l=2$ and arbitrary
branches $b$ is depicted in (d). The resulting RG-recursions often
permit an analytic continuation to arbitrary $b$ and/or $l$, for
a continuously defined dimension $d=1+\log_{l}b$ \cite{Li2016a}. }
\end{figure*}

We denote eigenvalues and normalized orthogonal eigenstates for $\mathcal{H}$
and $\mathcal{L}$ respectively as $\left\{ E_{i},\left|\psi_{i}\right\rangle \right\} $
and $\left\{ \lambda_{i},\left|\phi_{i}\right\rangle \right\} $ for
$0\leq i<N$. Note that the initial state $\left|s\right\rangle =\left|\phi_{0}\right\rangle $
is, in fact, the lowest eigenstate of the Laplacian with ${\cal L}\left|s\right\rangle =0$,
i.e., the associated eigenvalue is $\lambda_{0}=0$, while all other
Laplacian eigenvalues are positive. Ref. \cite{Childs04} has derived
a spectral function for ${\cal H}$ in terms of the Laplacian eigenstates
that is convenient for the discussion of translationally invariant
lattices:

\begin{eqnarray}
F\left(E\right) & =\left\langle w\left|\frac{1}{\gamma{\cal L}-E}\right|w\right\rangle =\sum_{i=0}^{N-1}\frac{\left|\left\langle w|\phi_{i}\right\rangle \right|^{2}}{\gamma\lambda_{i}-E}.\label{eq:function_f(E)}
\end{eqnarray}
The condition on the Hamiltonian eigenvalues,
\begin{equation}
F\left(E_{i}\right)=1,\label{eq:Fev}
\end{equation}
is provided in terms of the Laplacian eigenvalues. From the spectral
function, one can derive the overlap of any eigenstate with the initial
state as
\begin{equation}
\left|\left\langle s\left|\psi_{i}\right.\right\rangle \right|^{2}=\frac{1}{N\,E_{i}^{2}\,F^{\prime}\left(E_{i}\right)}.\label{eq:soverlap}
\end{equation}
The key objective of a quantum search concerns optimizing the transition
amplitude between the initial state and the target site, 
\begin{equation}
\left\langle w\left|e^{i{\cal H}t}\right|s\right\rangle =-\frac{1}{\sqrt{N}}\,\sum_{i=0}^{N-1}\frac{e^{iE_{i}t}}{E_{i}F^{\prime}\left(E_{i}\right)}.\label{eq:ws}
\end{equation}
The Hamiltonian ${\cal H}$ (i.e., $\gamma$) has to be optimized
such that this amplitude reaches a finite magnitude in the shortest
amount of time in the limit of large $N$. 

For regular lattices, the overlap of the eigenstates of the Laplacian
with any member of the site basis, in particular $\left|w\right\rangle $,
is uniform and \emph{independent} of $w$. However, this is generally
not true for fractals\textbf{ \cite{Rammal84}}; such heterogeneity
could lead to a large variability in the ``findability'' of a significant
number of ill-placed sites $w$ \cite{Phillip16}. We have to assume
(and will demonstrate below) that at typical sites $w$ of some fractal
networks, the overlaps with eigenvectors of the Laplacian still satisfy
\begin{equation}
\left|\left\langle w|\phi_{i}\right\rangle \right|^{2}\sim\frac{1}{N},\label{eq:up}
\end{equation}
as they are for the Fourier modes of the lattice. With that, and also
remembering that $\lambda_{0}=0$, Eq. (\ref{eq:function_f(E)}) can
be rewritten as

\begin{equation}
F(E)\sim-\frac{1}{N\thinspace E}+\frac{1}{\gamma}I_{1}+\frac{1}{N}\sum_{i=1}^{N-1}\frac{E}{\gamma\lambda_{i}\left(\gamma\lambda_{i}-E\right)},\label{eq:F_redux}
\end{equation}
defining the spectral $\zeta$-function \cite{Voros92,dunne2012KernelFractals}
\begin{equation}
I_{j}\sim\frac{1}{N}\,\sum_{i=1}^{N-1}\left(\frac{1}{\lambda_{i}}\right)^{j},\label{eq:DefIj}
\end{equation}
which will play a central role in the analysis. These quantities have
been considered before, in particular by Ref. \cite{Childs04}, to
examine search by CTQW on regular lattices, or in Ref. \cite{Agliari2011}
for fractals. In Ref. \cite{Agliari2011}, it was assumed that Eq.
(\ref{eq:DefIj}) requires complete knowledge of the entire Laplacian
spectrum, which is rarely achievable. Here, we want to point out that
$I_{j}$ can be reduced to the evaluation of the determinant of ${\cal L}$
and derivatives thereof. Using the fact that
\begin{eqnarray*}
\sum_{i=1}^{N-1}\ln\lambda_{i} & = & \ln\left[\frac{1}{\epsilon}\prod_{i=0}^{N-1}\left(\lambda_{i}+\epsilon\right)\right]_{\epsilon\to0},\\
 & = & \ln\left[\frac{1}{\epsilon}\det\left({\cal L}+\epsilon\right)\right]_{\epsilon\to0},
\end{eqnarray*}
 we have obtained in Ref. \cite{Li2016a} asymptotic behavior of the
spectral $\zeta$-function defined in Eq. (\ref{eq:DefIj}) as 
\begin{align}
I_{j} & \sim\left.\left(\frac{\partial}{\partial\epsilon}\right)^{j}\ln\left[\frac{1}{\epsilon}\det\left({\cal L}+\epsilon\right)\right]\right|_{\epsilon\to0}\label{eq:Nds}\\
 & \sim\begin{cases}
N^{\frac{2j}{d_{s}}-1}, & d_{s}<2j,\\
const, & d_{s}>2j,
\end{cases}
\end{align}
for fractal networks with the spectral dimension $d_{s}$. Thus, $d_{s}$
becomes the key characteristic for any network, such as those fractals
for which $d=d_{f}\not=d_{s}$, that determines whether the Grover
limit can be achieved. For example, as observed in Ref. \cite{Childs04},
this quantum search becomes optimal for lattices of any dimension
when there is a phase transition in the overlaps $\left|\left\langle s|\psi_{0}\right\rangle \right|^{2}$
and $\left|\left\langle s|\psi_{1}\right\rangle \right|^{2}$ , of
which the former rises while the latter declines for increasing $\gamma$.
This critical point occurs for 
\begin{equation}
\gamma\sim\gamma_{c}=I_{1}.\label{eq:gammacritical}
\end{equation}
Accordingly, we find for general fractal networks that 
\begin{equation}
\gamma_{c}\sim\begin{cases}
N^{\frac{2}{d_{s}}-1}, & d_{s}<2,\\
const, & d_{s}>2.
\end{cases}
\end{equation}

To obtain the runtime complexity for the quantum search, we have to
distinguish the following cases: For $d_{s}>4$, according to Eq.
(\ref{eq:Nds}), both $I_{1,2}$ remain constant. It is then self-consistent
to consider the spectral function in Eq. (\ref{eq:F_redux}) for energies
$\left|E\right|\ll\gamma_{c}\lambda_{1}$, which applies to both the
ground state $E_{0}$ and the first excited state $E_{1}$ of ${\cal H}$
near the optimal (``critical'') $\gamma_{c}$. Expanding the remaining
sum in Eq. (\ref{eq:F_redux}) to leading order in $E$ yields
\begin{equation}
F\left(E\right)\sim-\frac{1}{NE}+\frac{1}{\gamma}\,I_{1}+\frac{E}{\gamma^{2}}\,I_{2}+\ldots,\qquad\left(\left|E\right|\ll\gamma_{c}\lambda_{1}\right),\label{eq:FEexpand}
\end{equation}
 Since $F\left(E_{0,1}\right)=1$ from the eigenvalue condition in
Eq. (\ref{eq:Fev}), we obtain a consistent balance to leading and
sub-leading order only for $\gamma=\gamma_{c}=I_{1}$, thereby validating
Eq. (\ref{eq:gammacritical}), and for $\frac{1}{NE_{0,1}}\sim\frac{E_{0,1}}{\gamma^{2}}I_{2}\ll1$,
yielding
\begin{equation}
E_{0,1}\sim\pm\frac{1}{\sqrt{N}}\,\frac{I_{1}}{\sqrt{I_{2}}}=O\left(N^{-\frac{1}{2}}\right).\label{eq:E01}
\end{equation}
Then, the derivative of Eq. (\ref{eq:FEexpand}) provides $F^{\prime}\left(E_{0,1}\right)\sim2I_{2}\left/I_{1}^{2}\right.$
such that according to Eq. (\ref{eq:soverlap}) the initial state
$\left|s\right\rangle $ overlaps with equal and finite weight with
both, ground state and first excited state:

\begin{eqnarray}
\left|\left\langle s\mid\psi_{0,1}\right\rangle \right|^{2} & \sim\frac{1}{2} & .\label{eq:overlap-01}
\end{eqnarray}
As $E_{i}>\gamma_{c}\lambda_{1}$ for all $i\geq2$, higher energy
eigenstates do not contribute for large $N$, and we obtain from the
first two terms of the transition amplitude in Eq. (\ref{eq:ws}),
\begin{align}
\left|\left\langle w\left|e^{i{\cal H}t}\right|s\right\rangle \right|^{2} & \sim\frac{1}{N}\left|\frac{e^{iE_{0}t}}{E_{0}F^{\prime}\left(E_{0}\right)}+\frac{e^{iE_{1}t}}{E_{1}F^{\prime}\left(E_{1}\right)}\right|^{2},\\
 & \sim\frac{I_{1}^{2}}{I_{2}}\,\sin^{2}\left(\frac{2I_{1}}{\sqrt{I_{2}}}\,\frac{t}{\sqrt{N}}\right).\label{eq:ws_transition}
\end{align}
Thus, the transition probability oscillates and reaches its first
maximum at a time 
\begin{equation}
t=t_{{\rm opt}} \sim \frac{\sqrt{I_{2}}}{I_{1}}\,\sqrt{N}=O\left(N^{\frac{1}{2}}\right),\label{eq:topt}
\end{equation}
at which point the transition probability becomes
\begin{equation}
p_{{\rm opt}}=\left|\left\langle w\left|e^{i{\cal H}t_{{\rm opt}}}\right|s\right\rangle \right|^{2}\sim\frac{I_{1}^{2}}{I_{2}}=O(1).
\end{equation}
Finally, to find the targeted site $w$ with a probability of order
unity, we need to run the quantum search $\sim1/p_{{\rm opt}}$ times,
each for a time of $t_{{\rm opt}}$ at which a measurement must be
executed. Thus, the runtime complexity for a successful search is
given by
\begin{equation}
\frac{t_{{\rm opt}}}{p_{{\rm opt}}}\sim\left(\frac{I_{2}}{I_{1}^{2}}\right)^{\frac{3}{2}}\sqrt{N}=O\left(N^{\frac{1}{2}}\right),\qquad(d_{s}>4).\label{eq:complexity}
\end{equation}

For case $d_{s}=4$, $I_{1}$ remains constant while $I_{2}\sim\ln N$
acquires a logarithmic correction in the limit $d_{s}\to4$. With
that, the analysis of the previous case remains applicable, although
the condition $\left|E\right|\ll\gamma_{c}\lambda_{1}$ is merely
logarithmically satisfied. Thus, we obtain from Eq. (\ref{eq:complexity})
in this interpretation that 
\begin{equation}
\frac{t_{{\rm opt}}}{p_{{\rm opt}}}=O\left(N^{\frac{1}{2}}\ln^{\frac{3}{2}}N\right),\qquad(d_{s}=4).
\end{equation}

\begin{figure}
\centering
\begin{subfigure}{1.0\columnwidth}
\center
\includegraphics[width=0.75\columnwidth,angle=270]{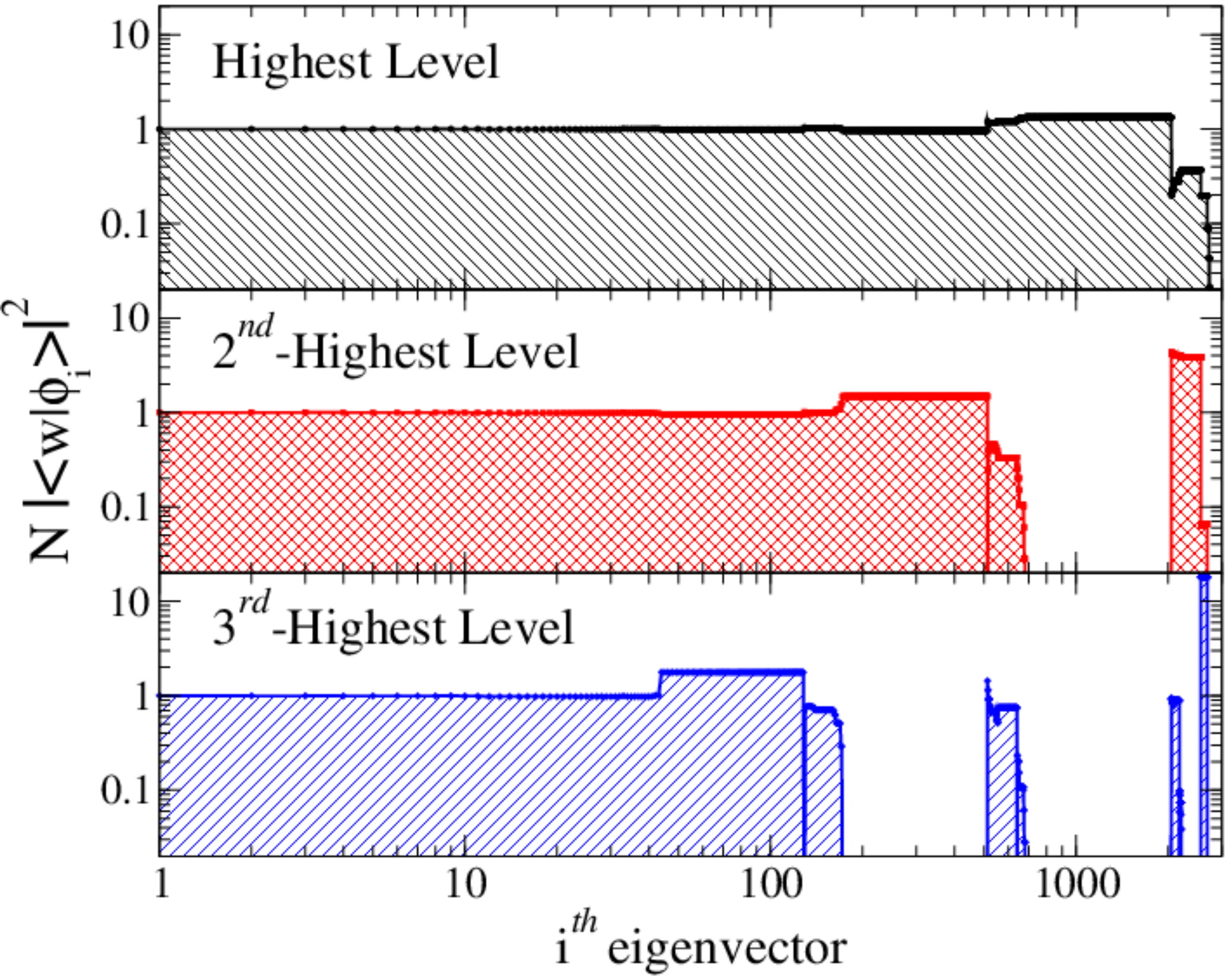}
\caption{}
\end{subfigure}
\begin{subfigure}{1.0\columnwidth}
\center
\includegraphics[width=0.75\columnwidth,angle=270]{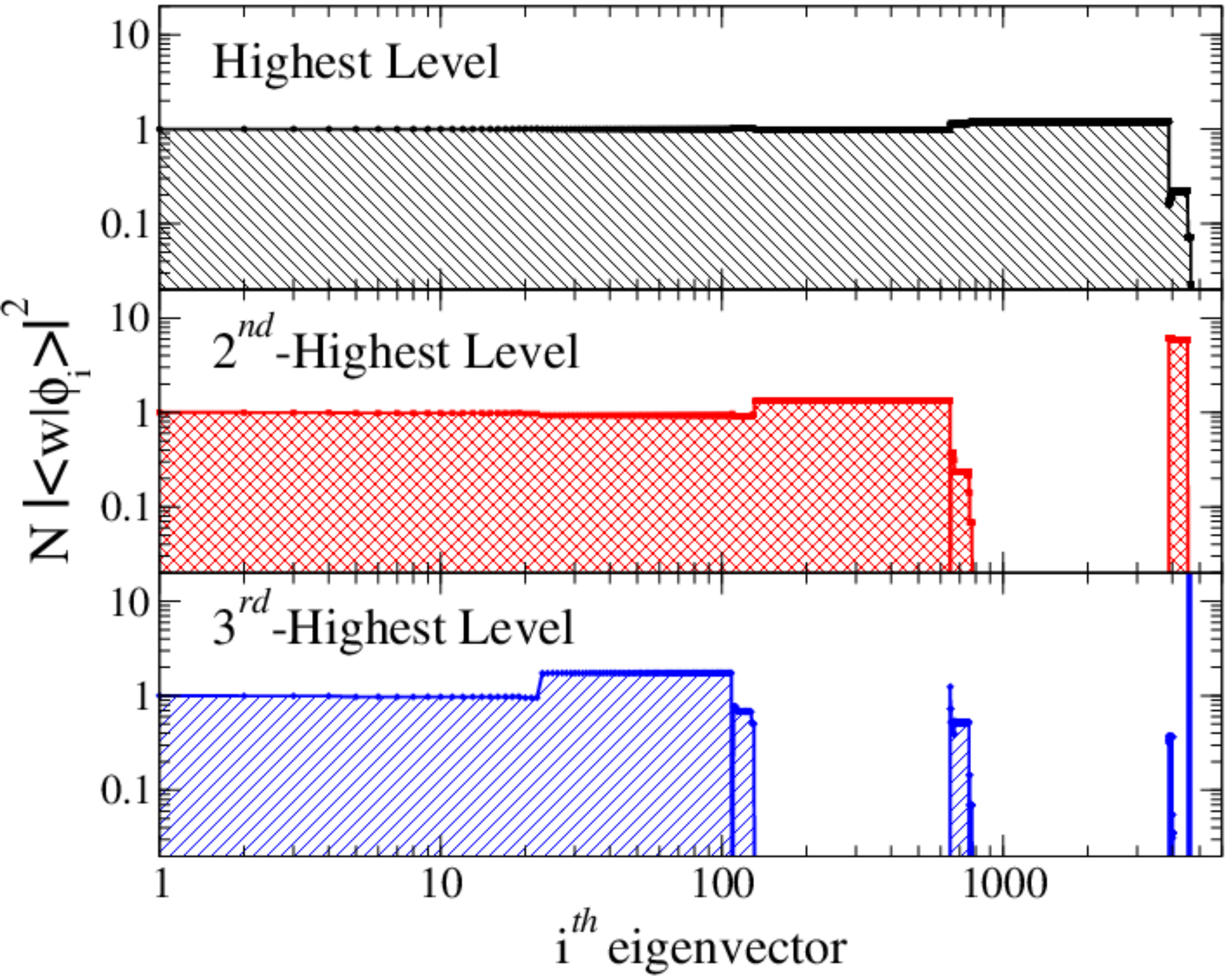}
\caption{}
\end{subfigure}
\captionsetup{justification=raggedright, singlelinecheck=false}
\noindent \caption{\label{fig:overlap_average-evec} Plot of the overlaps $\left|\left\langle w|\phi_{i}\right\rangle \right|^{2}$
in MKRG for the searched-for sites $w$ with the Laplacian eigenvectors
$\phi_{i}$ as a function of $i$, ordered such that respective eigenvalues
satisfy $\lambda_{i}\leq\lambda_{i^{\prime}}$ for any two indices
$i\leq i^{\prime}$. The MKRG used here \cite{MKpaper} rescales length
by $l=2$ with (a) $b=2$ and (b) $b=3$ branches in each RG-step
for an effective dimension $d=1+\log_{l}b$ of (a) $d=2$ and (b)
$d=2.585\ldots$. The RG has been iterated for $g=6$ generations
in the hierarchy in (a), forming a lattice of $N=2+\frac{b}{2b-1}\left[\left(2b\right)^{g}-1\right]=2732$
sites, and in (b) for $g=5$ with $N=4667$ sites. In the top panel
of both, (a) and (b), the overlaps (rescaled by a factor of $N$)
were averaged over all sites $w$ in the highest hierarchical level
$g$, in the middle panel overlaps were averaged only over those $w$
in level $g-1$, and in the respective bottom panel for level $g-2$.
Note that every level the number of sites increases by a factor of
$\sim2b$, such that the vast majority of all sites $w$ are typically
located in these highest levels of the hierarchy. For those, these
plots show that indeed $N\left|\left\langle w|\phi_{i}\right\rangle \right|^{2}\sim1$,
typically, as assumed in Eq. (\ref{eq:up}), although these overlaps
progressively vanish for those $w$ in lower levels for eigenvectors
of larger index $i$.}
\end{figure}

For case $2<d_{s}<4$, $I_{1}$ remains constant while $I_{2}\sim N^{\frac{4}{d_{s}}-1}$.
However, by Eq. (\ref{eq:E01}), this would imply $E_{0,1}\sim N^{-\frac{2}{d_{s}}}$
, which would violate the condition of $E_{0,1}\ll\gamma_{c}\lambda_{1}$
where $\lambda_{1}\sim\Lambda N^{-\frac{2}{d_{s}}}$. As a consequence,
the expansion in Eq. (\ref{eq:FEexpand}) is no longer is valid and
we have to reconsider Eq. (\ref{eq:F_redux}) anew at $\gamma=\gamma_{c}\sim I_{1}$,
but with $\gamma_{c}\lambda_{1}\sim E_{0,1}=e_{0,1}I_{1}\Lambda N^{-\frac{2}{d_{s}}}$.
Then, Eq. (\ref{eq:F_redux}) provides
\begin{equation}
F\left(E_{0,1}\right)\sim1-\frac{1}{I_{1}\Lambda e_{0,1}}N^{\frac{2}{d_{s}}-1}+\frac{e_{0,1}}{I_{1}\Lambda\left(1-e_{0,1}\right)}N^{\frac{2}{d_{s}}-1}+\ldots,
\end{equation}
where the two leading corrections cancel self-consistently with a
negative (positive) solution for $e_{0}$ ($e_{1}$). Then, $E_{0,1}F^{\prime}\left(E_{0,1}\right)\sim N^{\frac{2}{d_{s}}-1}$,
such that by Eq. (\ref{eq:ws_transition}), the transition probability
diminishes for falling $d_{s}$ and is at best 
\begin{eqnarray}
\left|\left\langle w\left|e^{i{\cal H}t}\right|s\right\rangle \right|^{2} & \apprle & \frac{1}{N}\left|\frac{1}{E_{0}F^{\prime}\left(E_{0}\right)}\right|^{2}\sim N^{1-\frac{4}{d_{s}}}.\label{eq:transbelow4}
\end{eqnarray}
 In turn, to accomplish any significant change in this transition
amplitude requires at time of at least $t_{opt}\apprge\left|\left\langle w\left|e^{i{\cal H}t}\right|s\right\rangle \right|\sqrt{N}\sim N^{1-\frac{2}{d_{s}}}$.
Thus, the runtime complexity finally is asymptotically bounded by
\begin{eqnarray}
\frac{t_{opt}}{p_{opt}} & \apprge & N^{\frac{2}{d_{s}}},\qquad\left(2<d_{s}<4\right).\label{eq:Db2T_runtime-complexity}
\end{eqnarray}
which for $d_{s}\to2$ also reproduces the known conclusion for the
\emph{2d} regular lattice, up to logarithmic corrections. 

Finally, we confirm numerically the assumption in Eq. (\ref{eq:up})
that for typical vertices $w$, the overlap with Laplacian eigenvectors
scales as $\left|\left\langle w|\phi_{i}\right\rangle \right|^{2}\sim1/N$.
For example, in Ref. \cite{Li2016a}, we have considered the Laplacians
for fractal networks in the Migdal-Kadanoff renormalization group
(MKRG) \cite{Migdal76,Kadanoff76}, which mimic the properties of
regular lattices quite closely and have $d_{s}=d_{f}=d$ but can take
one such values also for non-integer dimensions, as described in Fig.
\ref{fig: MKlattice}. Unlike for regular lattices, though, in MKRG
sites are arranged in a hierarchical network \cite{Berker79} in which
at each level of the hierarchy the system size expands by a factor
of $\sim2^{d-1}$. Those newly added sites all are locally equivalent,
but they are distinct from previous levels, making the network more
heterogeneous than the lattice it is meant to represent. However,
since they constitute by far the largest fraction and are the most
uniform, sites in the highest level of the hierarchy exhibit what
can be considered as the typical behavior. In Fig. \ref{fig:overlap_average-evec},
we have plotted the overlaps for searched-for sites $w$ with all
eigenvectors $\left|\phi_{i}\right\rangle $, $0\leq i<N$ of the
respective Laplacian but averaged separately over all $w$ in the
highest, $2^{{\rm nd}}$-highest, and $3^{{\rm rd}}$-highest levels
of the hierarchy for MKRG networks of $b=2$ after $g=6$ generations
of the hierarchy, and for $b=3$ after $g=5$ generation. For sites
$w$ in the highest level, the overlaps are essentially uniform and
satisfy Eq. (\ref{eq:up}) for all $i$. For $w$ on lower levels,
their overlaps with an increasing number of eigenvectors related to
the largest eigenvalues $\lambda_{i}$ outright vanishes, while the
non-vanishing overlaps remain with few exceptions uniform and $O\left(1/N\right)$.
This fact suggests that the runtime complexity differs mildly between
sites $w$ in different hierarchies. For our purpose here, we conclude
that sites $w$ in the highest level are most representative of the
behavior of any site on a regular lattice. 

In conclusion, we have studied search by a continuous-time quantum
walk on fractal networks and, by reference to properties of spectral
$\zeta$-functions \cite{Li2016a}, identified the dominant role of
the spectral dimension $d_{s}$ in controlling the search efficiency
and in setting the condition for attaining the Grover limit, for $d_{s}>4$.
Particularly, we reproduce the known results in regular lattices with
integer $d=d_{s}$ and generalize them to hyper-cubic lattices in
arbitrary dimensions $d$ using the Migdal-Kadanoff renormalization
group. Although this family of fractals is chosen to satisfy $d_{s}=d_{f}=d$,
the analysis in Ref. \cite{Li2016a} that leads to Eq. (\ref{eq:Nds})
implies the preeminence of $d_{s}$ also for search on other fractals,
as had been suggested previously in numerical studies \cite{Agliari2011}. 

\bibliographystyle{apsrev4-1}
\bibliography{Boettcher}

\end{document}